\documentclass[reprint, secnumarabic,amssymb, aps,nobibnotes,superscriptaddress, longbibliography,pre]{revtex4-2}

\usepackage{amsmath}               
\usepackage{amssymb}               
\usepackage{amsbsy}
\usepackage{amsfonts}
\usepackage{tabularx}
\usepackage{color}
\usepackage{bm}
\usepackage{graphics}
\usepackage{dcolumn}
\usepackage{gensymb}

\renewcommand{\vec}{\bm}
\newcommand{\kB}{k_\text{B}}

\newcommand{\wpl}{\omega_\text{p}}
\newcommand{\wc}{\omega_\text{c}}
\newcommand{\rL}{r_\text{L}}
\newcommand{\vth}{v_\text{th}}
\newcommand{\wuh}{\omega_\text{UH}}

\begin{document}
\title{Dynamic structure factor of the magnetized one-component plasma: crossover from weak to strong coupling}
\author{Hanno Kählert}
\author{Michael Bonitz}
\affiliation{Christian-Albrechts-Universit\"at zu Kiel, Institut f\"ur Theoretische Physik und Astrophysik, Leibnizstr. 15, 24098 Kiel, Germany}
\date{\today}

\begin{abstract}
Plasmas in strong magnetic fields have been mainly studied in two distinct limiting cases--that of weak and strong nonideality with very different physical properties. While the former is well described by the familiar theory of Braginskii, the latter regime is closer to the behavior of a Coulomb liquid. Here we study in detail the transition between both regimes. We focus on the evolution of the dynamic structure factor of the magnetized one-component plasma from weak to strong coupling, which is studied with first-principle molecular dynamics simulations. The simulations show the vanishing of Bernstein modes and the emergence of higher harmonics of the upper hybrid mode across the magnetic field, a redistribution of spectral power between the two main collective modes under oblique angles, and a suppression of plasmon damping along the magnetic field. Comparison with results from various models, including the random phase approximation, a Mermin-type dielectric function, and the Quasi-Localized Charge Approximation show that none of the theories is capable of reproducing the crossover that occurs when the coupling parameter is on the order of unity. The findings are relevant to the scattering spectra, stopping power, and transport coefficients of correlated magnetized plasmas.
\end{abstract}
\maketitle

Strong particle interactions affect the thermodynamic, transport, and dielectric properties of plasmas. Conditions required for these effects to occur are found in very dense or cold systems, or when highly charged particles are involved. More specifically, in classical plasmas, the condition for the Coulomb coupling parameter,
\begin{align}
    \Gamma=\frac{Q^2}{4\pi \epsilon_0\,a\,\kB T},
\end{align}
where $Q$ is the charge, $T$ the temperature, and $a=[3/(4\pi n)]^{1/3}$ the Wigner-Seitz radius (density $n$), is $\Gamma\sim\mathcal O(1)$. Examples range from ions in warm dense matter~\cite{graziani2014book, bonitz2020} (solid densities), ultra cold neutral~\cite{lyon2016ultracold} and non-neutral plasmas~\cite{dubinreview} (mK temperatures), to complex (or dusty) plasmas~\cite{bonitz-introcomplex} (highly charged dust particles). Recent experimental advances in the magnetic confinement of ultra cold neutral plasmas~\cite{gorman2021prl}, high energy density matter~\cite{santos2018pop}, and dusty plasmas~\cite{thomas2016pop,kaehlert2012prl,bonitz2013,hartmann2013}, as well as theoretical efforts concerning, e.g., the stopping power~\cite{bernstein2020pre,jose2020pop, jose2021pop,bernstein2021pop} and transport coefficients~\cite{ott2011prl,feng2014pre,dzhumagulova2014pre,ott2015pre,dzhumagulova2016pre,ott2017pre,baalrud2017pre,schreiner2020pre, vidal2021pop} demonstrate growing interest in the physics of \textit{magnetized} strongly correlated plasmas--conditions relevant to the outer layers of neutron stars~\cite{harding2006rpp,chabrier2006, baiko2009pre,baiko2010jpcs, Potekhin2015ssr}, confined antimatter~\cite{danielson2015rmp,fajans2020pop}, or
magnetized target fusion~\cite{gomez2014prl,schmit2014prl}. In this challenging regime, the familiar theory of Braginskii~\cite{braginskii1965} is no longer applicable, and new theoretical concepts as well as first-principle simulations are required.

The dynamic structure factor (DSF) is highly relevant for both experiment and theory as it determines the x-ray Thomson scattering signal, a diagnostic used in dense plasma and warm dense matter experiments~\cite{glenzer2007prl,glenzer2009,kraus2016pre,kraus2018,bott2019pre}, and provides access to thermodynamic, transport~\cite{hansen2006, kaehlert2020prr,zeng2021prr}, and dielectric properties~\cite{magyar2021pre}, including the wave spectra~\cite{mithen2012pre,arkhipov2017prl,hamann2020prb,hamann2020cpp, magyar2021pre}. While waves in weakly coupled ($\Gamma\ll 1$) magnetized plasmas have been investigated theoretically in the seminal work of Bernstein~\cite{bernstein1958pr} and are documented in textbooks~\cite{bellan2008book,alexandrov2013}, molecular dynamics (MD) simulations, the Quasi-Localized Charge Approximation (QLCA)~\cite{golden2000}, and harmonic lattice theory have provided insight into the collective modes in very strongly coupled ($\Gamma\gtrsim 10$) magnetized two~\cite{uchida2004prl,jiang2007pop, hou2009pop, bonitz2010prl, ott2011pre} and three-dimensional systems~\cite{ott2012,yang2012pop,kaehlert2013, ott2013}. These conditions are encountered, e.g., in strongly coupled rotating dusty plasmas, where the Coriolis force in a rotating reference frame has been used to emulate the Lorentz force on charged particles in a magnetic field~\cite{kaehlert2012prl,bonitz2013, hartmann2013}. A common feature found in both coupling regimes is the appearance of several high frequency modes, including the famous Bernstein modes~\cite{bernstein1958pr}. The intermediate regime with $\Gamma\sim 1$, and the transition from weak to strong coupling, however, are largely unexplored, yet directly relevant, e.g., to the evolving field of magnetically confined ultra cold neutral plasmas~\cite{gorman2021prl}.

Therefore, in this work, first-principle MD simulations are conducted to explore the DSF and the wave spectra \textit{across} coupling regimes, including the moderately coupled regime, where a theoretical description is particularly challenging. As the coupling is increased from a weakly correlated state, the simulations show (i) the damping and finally the vanishing of Bernstein modes, (ii) the formation of higher harmonics of the upper hybrid mode, (iii) a redistribution of spectral power between the two principal collective modes under oblique angles, and (iv) a suppression of plasmon damping along the magnetic field. To connect with theory, the results are compared with the random phase approximation (RPA), a Mermin-type extension including collisions, and the QLCA. None of the theories provides a consistent description of the observations as the coupling regimes are crossed.

This work is organized as follows. The molecular dynamics simulations and the theoretical methods are briefly discussed in Sec.~\ref{sec:simmethod} and Sec.~\ref{sec:theory}, respectively. Results for the DSF are then presented in Sec.~\ref{sec:results}. The findings are discussed and summarized in Sec.~\ref{sec:conclusions}. Explicit expressions for the low order moments of the DSF are provided in the Appendix.

\section{Simulation Method}\label{sec:simmethod}
The system we study is the magnetized one-component plasma, where a single charged species is embedded in a uniform neutralizing background of the opposite charge and subject to the Lorentz force induced by an external magnetic field. The simulations have been performed using the LAMMPS code~\cite{plimpton1995jcp}, with an integration scheme adapted to strong magnetic fields~\cite{spreiter1999}. Computation of the DSF,
\begin{align}
S(\vec k,\omega)=\frac{1}{2\pi\,N}\int_{-\infty}^\infty \langle n(\vec k,t)n(-\vec k,0)\rangle e^{i\omega t}\,dt,
\end{align} 
where $n(\vec k,t)=\sum_i e^{-i\vec k\cdot \vec r_i(t)}$ is the Fourier transformed particle density, proceeds as in previous work~\cite{kaehlert2019,kaehlert2020prr}. The timestep $\Delta t$ was chosen as $\Delta t\, \wpl=0.0015$ for the lowest coupling strength, $\Gamma=0.125$, and was increased to $\Delta t\, \wpl=0.003$ ($\Delta t\, \wpl=0.01$) for $\Gamma=0.25$ and $\Gamma=0.5$ ($\Gamma\ge 1$). Here, $\wpl=[Q^2 n/(\epsilon_0 m)]^{1/2}$ is the plasma frequency. Most of the simulations have been carried out with $N=80,000$ particles but for some simulations with $\Gamma\ge 1$ smaller system sizes have been used as well ($N=10,000$).

After an initial equilibration period without external field, a magnetic field $\vec B=B\,\hat{\vec e}_z$ along the $z$-axis of the coordinate system is turned on, and the particle density is recorded for various wave vectors $\vec k$ along (component $k_\parallel$) and across $\vec B$ (component $k_\perp$), as well as for oblique angles $\theta=\angle (\vec k,\vec B)$. For future reference, the magnetization will be given as 
\begin{align}
    \beta=\frac{\wc}{\wpl},
\end{align}
the ratio of the cyclotron frequency, $\wc=QB/m$, and the plasma frequency, $\wpl$.

\section{Theory}\label{sec:theory}
For a direct comparison with theory, the fluctuation-dissipation theorem is invoked~\cite{kubo1966rpp}, which connects the DSF with the inverse of the longitudinal dielectric function, $\epsilon(\vec k,\omega)$, according to
\begin{align}
    S(\vec k,\omega)= -\frac{\kB \,T}{\pi\,\omega\, n\,\hat v_c(k)} \text{Im}\left[\epsilon^{-1}(\vec k,\omega)\right],
\end{align}
where $\hat v_c(k)=Q^2/(\epsilon_0\, k^2)$ is the Fourier transform of the Coulomb potential. For the dielectric function, the result from the Vlasov equation (or random phase approximation, RPA)~\cite{ichimaru1973book, alexandrov2013},
\begin{align}\label{eqn:RPA}
\epsilon_\text{RPA}&(\vec k,\omega)=\\
&1+\frac{1}{k^2\lambda^2}\left[ 1+\sum_{n=-\infty}^\infty   \frac{\omega}{\omega-n\,\omega_\text{c}} I_n(\eta)e^{-\eta}\, \zeta_n \,Z(\zeta_n)\right],\nonumber
\end{align}
which is a mean-field theory without collisional effects, and a Mermin-type extension with a particle number (but neither momentum nor energy) conserving Bhatnagar-Gross-Krook (BGK) collision operator (relaxation rate $\nu$) are employed~\cite{nersisyan2011,alexandrov2013},
\begin{align}\label{eqn:BGK}
 &\epsilon_\text{BGK}(\vec k,\omega) = \nonumber \\
 &1+ \frac{(\omega+i\nu)[\epsilon_\text{RPA}(\vec k,\omega+i\nu)-1]}{\omega+i\nu[\epsilon_\text{RPA}(\vec k,\omega+i\nu)-1]/[\epsilon_\text{RPA}(\vec k,0)-1]}.
\end{align}
The arguments of the modified Bessel function $I_n(\eta)$ and the plasma dispersion function $Z(\zeta_n)$ read $\eta=k_\perp^2 \rL^2$ and $\zeta_n=(\omega-n\,\wc)/(\sqrt{2}|k_\parallel|v_\text{th})$. Here, $\vth=(\kB T/m)^{1/2}$ is the thermal velocity, $\lambda=\vth/\wpl$ the Debye length, and $\rL=\vth/\wc$ the Larmor radius.

The roots of the dielectric function, $\epsilon(\vec k,\omega)=0$ , yield the spectrum of longitudinal modes. The spectrum perpendicular to the field ($\vec k\perp \vec B$) consists of a set of Bernstein modes near the harmonics of the cyclotron frequency~\cite{bellan2008book}, all of which are strictly real in the RPA. For one of the modes, the upper hybrid frequency, $\wuh=\sqrt{\wpl^2+\wc^2}$, is the long wavelength limit while all other modes start at harmonics of the cyclotron frequency. Parallel to the field ($\vec k\parallel \vec B$), one encounters the plasmon, which is not affected by the magnetic field and identical to the usual plasmon mode in the unmagnetized OCP. The BGK dielectric function~\cite{nersisyan2011} adds the effect of collisional damping, which is missing in the RPA.

\begin{figure}
\includegraphics{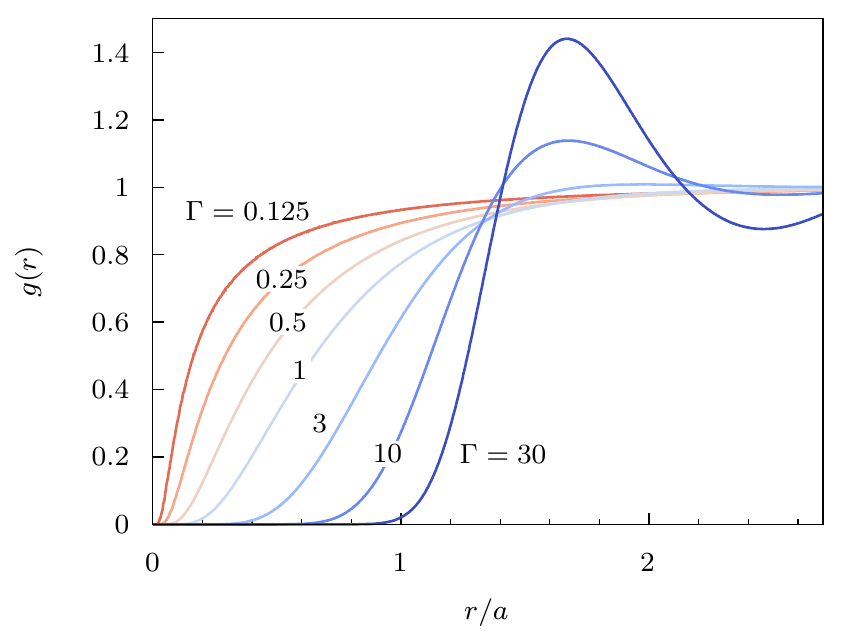}
\caption{\label{fig:pdfs}Pair distribution function $g(r)$ of the OCP for various values of $\Gamma$, as indicated in the figure. Note that $g(r)$ is independent of the magnetic field.}
\end{figure}

The above approaches are expected to be applicable in the weakly correlated regime, but do not extend into the regime of strong coupling. For very strongly coupled systems, where particles are trapped in local potential minima of the potential energy surface, the QLCA of Kalman and Golden~\cite{kalman1990,golden2000} has been successfully applied to the magnetized OCP~\cite{ott2012,kaehlert2013,ott2013}. The key quantity in the QLCA is the dynamical matrix $D_{\alpha\beta}(\vec k)=L_{\alpha\beta}D_L(k)+T_{\alpha\beta}D_T(k)$, where $L_{\alpha\beta}=k_\alpha k_\beta/k^2$ ($T_{\alpha\beta}=\delta_{\alpha\beta}-L_{\alpha\beta}$) is the longitudinal (transverse) projection operator and $D_{L/T}(k)$ the associated components of the dynamical matrix ($\alpha,\beta\in\{x,y,z\}$)~\cite{golden2000}, given by
\begin{align}\label{eqn:DT}
\frac{D_T(k)}{\wpl^2}=\int_0^\infty \frac{h(r)}{r}\left[\frac{\sin(kr)}{kr}+3\frac{\cos(kr)}{(kr)^2}-3\frac{\sin(kr)}{(kr)^3}\right]dr
\end{align}
and $D_L(k)=-2D_T(k)$. The latter require the pair distribution function $g(r)=h(r)+1$ of the plasma as input, which is available from the simulations, see Fig.~\ref{fig:pdfs}. The wave spectrum is then computed from~\cite{ott2012,kaehlert2013,ott2013}
\begin{align}\label{eqn:QLCA}
    \left(\omega^2\delta_{\alpha\beta} +i\omega\wc \sigma_{\alpha\beta} -\wpl^2 L_{\alpha\beta}-D_{\alpha\beta} \right)q_\beta=0,
\end{align}
where $\sigma_{\alpha\beta}$ accounts for the Lorentz force and has the only non-zero components $\sigma_{yx}=-\sigma_{xy}=1$, $\omega$ denotes the mode frequency and $q_\beta(\vec k,\omega)$ the components of the displacement vector.

The dispersion and polarization properties of the three collective modes that result from Eq.~\eqref{eqn:QLCA} have been studied in detail previously~\cite{ott2012,kaehlert2013,ott2013}. For the dynamic structure factor, which is directly coupled to the longitudinal current, modes with a longitudinal component are particularly relevant. Similar to the RPA above, for $\vec k\perp\vec B$ ($\vec k \parallel \vec B$), the dominant collective mode is the upper hybrid (plasmon) mode, both of which are now modified by correlations in the QLCA. As in the RPA, the plasmon dispersion is independent of the magnetic field.

The phenomena discussed in the following occur on very different time and length scales, depending on $\Gamma$ and $\beta$. Relevant physical parameters include the Debye screening length $\lambda$ (in particular for $\Gamma \lesssim 1$) and the Larmor radius $\rL$. In the context of transport properties, the classical distance of closest approach, $b=Q^2/(4\pi \epsilon_0 \kB T)$, and the mean free path between collisions, $\lambda_\text{mfp}$, have additionally been used to define magnetization regimes,  depending on the ratios of $\rL$ and $\{\lambda_\text{mfp},\lambda,b,a\}$~\cite{braginskii1965,baalrud2017pre}. Note that, in the strongly coupled regime, $\Gamma\gtrsim 1$, the Wigner-Seitz radius $a$ eventually becomes smaller than the distance of closest approach and larger than the Debye screening length. Two useful relations between the length scales are
\begin{align}
     &\frac{\lambda}{a}=\frac{1}{\sqrt{3\,\Gamma}},
 &\frac{\rL}{a}=\frac{1}{\sqrt{3\,\Gamma}\beta}.
 \end{align}

For the discussion of the DSF and the wave spectra, time scales play an equally important role. In unmagnetized plasmas, longitudinal plasma waves occur near the plasma frequency $\wpl$. On the other hand, the inverse of the cyclotron frequency $\wc$ is the natural time scale for charged particles in a magnetic field. Thus, the parameter $\beta=\wc/\wpl$ as defined above should provide a useful estimate for the importance of magnetization effects on the wave spectra. In addition, the collision frequency $\nu_\text{c}=\vth / \lambda_\text{mfp}$ may be compared with the cyclotron frequency to estimate the effect of collisions on the gyromotion. This yields $\nu_\text{c}/\omega_\text{c}=\rL/\lambda_\text{mfp}$. In the context of the BGK dielectric function [Eq.~\eqref{eqn:BGK}], we use $\nu/\wc$ instead, where $\nu$ is the relaxation rate for the one-particle distribution function.

The low order moments of the DSF are known exactly and can be related to the static properties of the OCP. Explicit expressions for the moments are summarized in Appendix~\ref{sec:appendix}.

\section{Results}\label{sec:results}
In the following, the DSF of the magnetized OCP will be investigated for a wide range of coupling strengths, from a weakly correlated state with $\Gamma=0.125$ up to a strongly coupled system with $\Gamma=30$. The evolution of the pair distribution function across this regime is illustrated in Fig.~\ref{fig:pdfs}. As the coupling increases, the correlation hole becomes larger and, around $\Gamma\approx 1-3$, a maximum develops at $r/a\approx 1.7$, indicating the onset of short range order. As will be seen below, the coupling strength also manifests itself in \textit{dynamical} features, which are directly observable in the DSF.

\subsection{Wave vector (quasi) perpendicular to the magnetic field}
\subsubsection{Existence and vanishing of higher harmonics}\label{sssec:HH}
\begin{figure}
\includegraphics{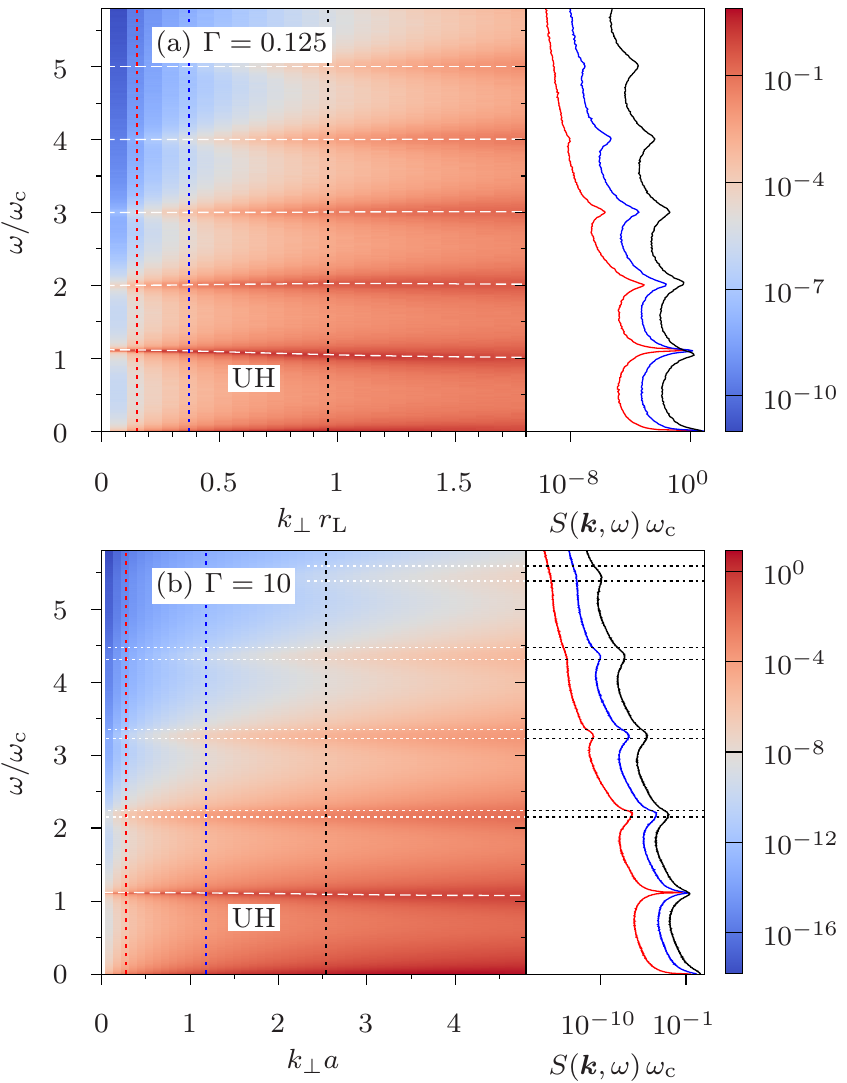}
\caption{\label{fig:dispersion}DSF for $\vec k\perp \vec B$ with $\beta=2$ in (a) the weakly coupled and (b) the strongly coupled regime. The color scale shows values of $S(\vec k,\omega)\,\wc$. The dashed lines depict (a) the dispersion relation from the RPA and (b) the QLCA solution for the upper hybrid mode. The short dashed lines in (b) correspond to harmonics of $\wuh$ and $\omega_\infty$, see the text for details. The side panels show cuts for fixed wave numbers, which are indicated by the dashed vertical lines. Note that the wave numbers on the horizontal axis are scaled differently in (a) and (b).}
\end{figure}

The DSF for a strongly magnetized ($\beta=2$) weakly coupled ($\Gamma=0.125$) plasma with wave vectors perpendicular to the external field is presented in Fig.~\ref{fig:dispersion}(a). It shows several peaks that can be attributed to Bernstein modes close to the cyclotron harmonics. The lowest of the modes begins at the upper hybrid frequency in the $k_\perp\to 0$ limit. While their positions are in excellent agreement with the RPA dispersion relation, obtained from the roots of $\epsilon_\text{RPA}(\vec k,\omega)$, an important difference occurs at small wave numbers, where the high order peaks disappear below a critical wave number. This is shown in more detail in the side panel, where cuts for fixed wave numbers are displayed. As the wave number is decreased (from right to left), the intensity of the high frequency peaks becomes weaker until they eventually vanish. Such behavior is similar to the vanishing of the higher harmonics in \textit{strongly coupled} magnetized plasmas, see Fig.~\ref{fig:dispersion}(b). However, in this regime, the high order modes now appear in close vicinity of the harmonics of the upper hybrid (UH) mode. The UH mode is in excellent agreement with the QLCA dispersion relation. The peak frequencies of the higher harmonics are located roughly between the $k_\perp\to 0$ and $k_\perp\to\infty$ limits of the QLCA, see the side panel. The latter is given by $\omega_\infty=\frac{1}{2}\left[\sqrt{\wc^2+4\,\omega_\text{E}^2}+|\wc|\right]$, where $\omega_\text{E}=\wpl/\sqrt{3}$ is the Einstein frequency of the OCP~\cite{ott2012,kaehlert2013}. In case of the high order modes and large $k_\perp$, the peaks tend to be closer to $\omega_\infty$ while the second harmonic at small $k_\perp$ is closer to $2\,\wuh$. Similar behavior was observed previously in magnetized two-dimensional Yukawa liquids~\cite{bonitz2010prl,ott2011pre} and in the current spectra of the magnetized OCP~\cite{ott2012}.

\begin{figure}
\includegraphics{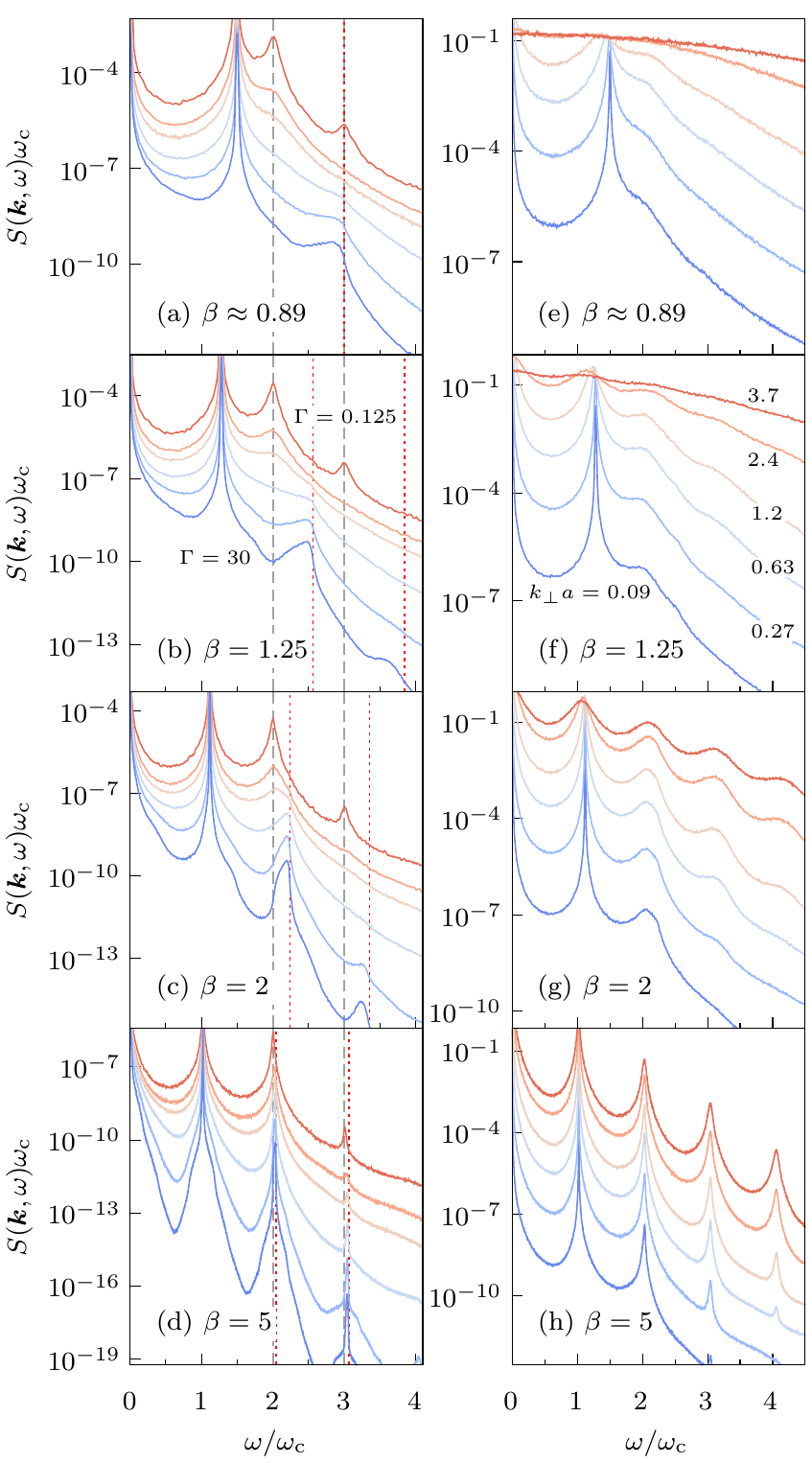}
\caption{\label{fig:transition}Transition of the DSF from weak to strong coupling for $\vec k\perp \vec B$. (a)--(d) $k_\perp a\approx 0.0905$ for coupling parameters $\Gamma\in\{0.125,\, 0.5,\, 1,\, 3,\, 10,\, 30\}$ (from top to bottom). Short dashed (long dashed) vertical lines show harmonics of the upper hybrid (cyclotron) frequency. (e)--(h) DSF for $\Gamma=1$ at wave numbers $k_\perp a\in \{ 0.0905, 0.271, 0.633, 1.18, 2.35, 3.71\}$ (from bottom to top).}
\end{figure}

The transition between the weak and strong coupling regime is illustrated in Fig.~\ref{fig:transition}. The left column depicts the DSF at the smallest $k_\perp a$ available in the simulations. Consider first Fig.~\ref{fig:transition}(a) with $\beta=2/\sqrt{5}\approx 0.89$. In the most weakly coupled system, $\Gamma=0.125$, there are two clear maxima at the $n=2$ and $n=3$ cyclotron harmonics ($\omega=n\,\wc$). As the coupling increases to $\Gamma=0.5$, the $n=3$ peak disappears, and the second harmonic becomes very weak. It vanishes for $\Gamma=3$, where a weak shoulder emerges at the second harmonic of the upper hybrid frequency, which, in this particular case, is identical to the third cyclotron harmonic. The shoulder like structure becomes more pronounced at higher coupling. At $\Gamma=30$, a local maximum is formed just below $\omega=2\,\wuh=3\,\wc$.

For $\beta=1.25$ [Fig.~\ref{fig:transition}(b)] the harmonics of $\wc$ and $\wuh$ are more clearly separated. Analogous to the previous case, the cyclotron harmonics disappear at intermediate coupling strengths but it now becomes apparent that the shoulder in the strongly coupled systems emerges just below the second harmonic of $\wuh$, where the DSF first increases and then starts to fall off rapidly. This feature becomes even more pronounced at $\beta=2$ [Fig.~\ref{fig:transition}(c)], where the peak between $2\,\wc$ and $2\,\wuh$ never vanishes at any of the $\Gamma$ considered, yet develops a pronounced asymmetry. It does disappear, however, in case of the third harmonic, i.e., in the interval $(3\,\wc,3\,\wuh)$. At the strongest magnetization, $\beta=5$ [Fig.~\ref{fig:transition}(d)], the line shape changes qualitatively and closely resembles the current spectra in previous simulations at $\Gamma=100$~\cite{ott2012}. A distinction between harmonics of $\wc$ and $\wuh$ becomes increasingly difficult because $\wc\approx \wuh$ for $\beta\gg 1$.

The right column of Fig.~\ref{fig:transition} shows the DSF at $\Gamma=1$ for a range of wave numbers. It is apparent that $\beta\gtrsim 1$ is required to observe the high frequency modes in a plasma with intermediate coupling strengths, even for larger wave numbers, where their intensity is higher, see also Fig.~\ref{fig:dispersion}.

\begin{figure}
\includegraphics{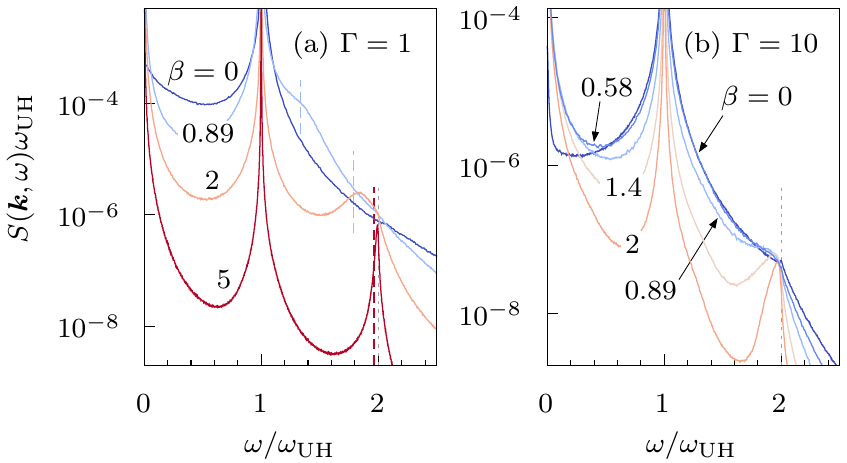}
\caption{\label{fig:posharmonic}DSF for $\vec k\perp \vec B$ at (a) $\Gamma=1$ and (b) $\Gamma=10$. The wave number is $k_\perp a=0.181$. The long dashed vertical lines in (a) show the position of the second cyclotron harmonic at $2\,\wc$. The short dashed lines indicate the second harmonic of the upper hybrid frequency. The magnetization is given in the figure. Note that for $\beta\to 0$ we have $\wuh\to\wpl$.}
\end{figure}

The different nature of the second harmonics in moderately and strongly coupled plasmas is exemplified further in Fig.~\ref{fig:posharmonic}, where frequencies have been normalized by the upper hybrid frequency, and the limit of small magnetization is considered. The position of the second harmonic for $\Gamma=1$ shifts as $\beta$ is varied [Fig.~\ref{fig:posharmonic}(a)]. It is well described by twice the cyclotron frequency, which suggests that this mode is a remnant of the Bernstein modes. In the strongly coupled system with $\Gamma=10$, however, the peak position (for large $\beta$) or the rapid decay of the DSF (for small $\beta)$ are always found in the close vicinity of $2\,\wuh$ [Fig.~\ref{fig:posharmonic}(b)]. Even in the unmagnetized limit, $\beta=0$, this feature persists. It has been observed previously in MD simulations of the OCP~\cite{hartmann2009, korolov2015cpp}, where it was considered an indication for the excitation of the second plasmon harmonic. The strong similarity with the upper hybrid mode suggests that the mechanisms for the excitation of harmonics in magnetized and unmagnetized plasma could be closely related and that the effect becomes strongly amplified as the field strength increases. It would be interesting to study these mechanisms in more strongly coupled systems, where higher order harmonics can be observed even in unmagnetized systems, but this is beyond the scope of this work.

\begin{figure}
\includegraphics{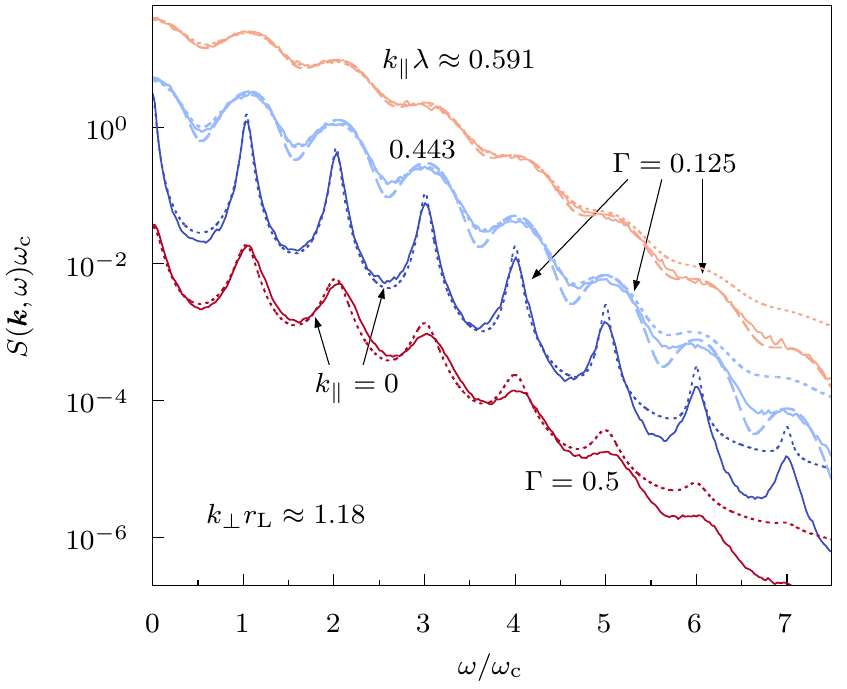}
\caption{\label{fig:dampbernstein}DSF for $\beta=2$ and $k_\perp r_\text{L}\approx 1.18$. The coupling parameter is $\Gamma=0.125$ for the three upper curves and $\Gamma=0.5$ for the lowest curve. The two uppermost curves have a finite wave number parallel to the magnetic field, as indicated in the figure. The corresponding angles between $\vec k$ and $\vec B$ are $\theta\approx 76\degree$ and $\theta\approx 79\degree$. The long dashed (short dashed) lines depict the RPA (BGK) DSF. For the BGK data, the relaxation rate is $\nu/\wc=0.05$ ($\Gamma=0.125)$ and $\nu/\wc=0.14$ ($\Gamma=0.5)$. The data have been shifted for clarity.}
\end{figure}

Returning to the high frequency spectrum in weakly to moderately coupled systems, the damping of Bernstein modes is studied in more detail in Fig.~\ref{fig:dampbernstein}. For strictly $\vec k \perp \vec B$, the RPA dispersion relation is purely real, i.e., modes propagate without damping. The simulations, however, show a pronounced broadening of the various peaks. This can be explained qualitatively with Eq.~\eqref{eqn:BGK}, which contains collisional damping via a BGK type collision operator. With a proper choice of the relaxation rate $\nu$~\footnote{Note that the relaxation rate was chosen empirically and was not fitted to the simulation data.}, the line shape of the simulations can be well reproduced up to $\omega\approx 5\,\wc$ for $\Gamma=0.125$. A larger $\nu$ is required to capture the more strongly coupled system with $\Gamma=0.5$, where the individual peaks have become significantly less pronounced, for the same $k_\perp \rL$. Moreover, the deviations in the high frequency tail become stronger, and the peak positions show slight deviations from the simulation data. 

The two uppermost curves illustrate the damping effect of a small wave vector component parallel to the magnetic field. Here, the RPA result is also shown, which now contains cyclotron damping via the plasma dispersion function $Z(\zeta_n)$ [see Eq.~\eqref{eqn:RPA}] but still misses collisional damping. While the RPA underestimates the peak broadening, especially in the low frequency domain, it provides a much improved agreement at high frequencies compared to the DSF with collisions, suggesting that the simple BGK type collision operator with a constant relaxation rate cannot capture the dynamics across the entire frequency range.

\subsubsection{Waves in weakly magnetized plasmas}
The focus so far was mainly on plasmas with strong magnetization, where all Bernstein modes are located above the upper hybrid frequency. According to the RPA dispersion relation, Bernstein modes below the upper hybrid frequency start at $(n+1)\,\wc$ at $k_\perp=0$ and approach $n\,\wc$ for $k_\perp\to\infty$. This is in contrast to Bernstein modes above $\wuh$, which have the same limit for small and large wave numbers, cf. Fig.~\ref{fig:dispersion}. 

\begin{figure}
\includegraphics{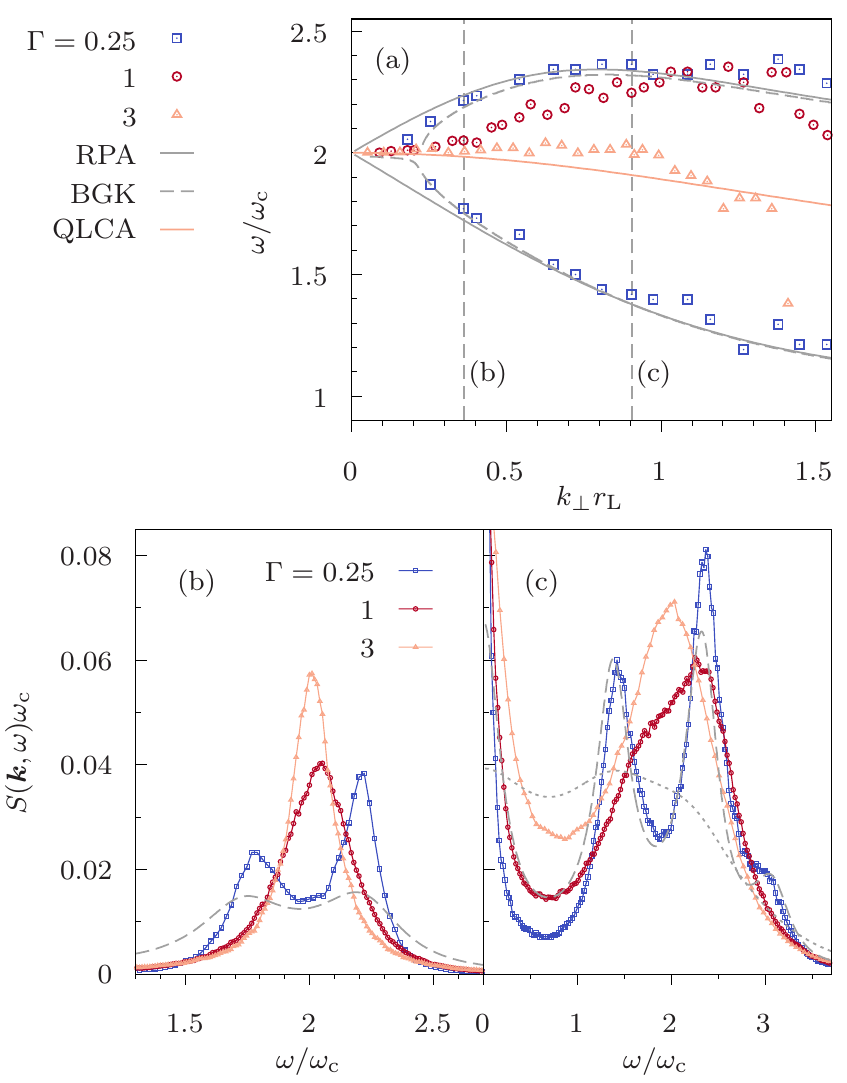}
\caption{\label{fig:dispb058}(a) Peak position from the DSF with a magnetization $\beta=1/\sqrt{3}$ ($\wuh/\wc =2$) for different $\Gamma$ as a function of $k_\perp r_\text{L}$. Also depicted is the dispersion relation from the RPA and QLCA ($\Gamma=3$) as well as the peak position from the BGK DSF with $\nu/\wpl=0.15$. Panels (b) and (c) show the DSF for the two wave numbers indicated by the dashed vertical lines in (a). The dashed lines show the BGK DSF with $\nu/\wpl=0.15$. The short dashed line in (c) additionally depicts the BGK result with $\nu/\wpl=0.5$. Note that the $k_\perp \rL$ used for $\Gamma=3$ deviates slightly ($\sim 1\%$) from the corresponding values for $\Gamma=0.25$ and $\Gamma=1$.}
\end{figure}

Figure~\ref{fig:dispb058} shows the DSF for a system with $\beta=1/\sqrt{3}\approx 0.58$, where the upper hybrid frequency and the second harmonic of the cyclotron frequency coincide ($2\,\wc=\wuh$). Since the $k_\perp\to 0$ limits of the two lowest RPA modes are the same, this constitutes a particularly challenging situation. For the weakly coupled system with $\Gamma=0.25$, the simulations show a single peak at the smallest wave number, which splits into two as $k_\perp$ increases. The peak positions are in very good agreement with the RPA dispersion relation, except for the smallest wave numbers, see Fig.~\ref{fig:dispb058}(a). Increasing the coupling to $\Gamma=1$, only a single peak remains. Its position is initially between the two RPA modes and appears to follow the upper mode beyond $k_\perp \rL\approx 1$, see also Fig.~\ref{fig:dispb058}(b) and Fig.~\ref{fig:dispb058}(c), which depict the DSF in the relevant frequency range. A possible remnant of the lower peak remains visible in Fig.~\ref{fig:dispb058}(c). At $\Gamma=3$, the peak position is almost constant up to $k_\perp \rL\approx 1$ and then decreases, which is accompanied by a pronounced peak broadening. 

Note that, in general, the peak positions from the DSF do not necessarily provide a good estimate for (the real part of) the dispersion relation of the collective modes, especially when the peaks have a large width or when two modes overlap, see also the discussion in Ref.~\cite{hamann2020cpp}. The dashed line in Fig.~\ref{fig:dispb058}(a) shows an attempt to explain the above observations with Eq.~\eqref{eqn:BGK}. It shows the peak positions of the DSF within the BGK model with $\nu/\wpl=0.15$, where the peaks now posses a finite width. On the one hand, the slight increase (decrease) of the frequency of the lower (upper) mode, compared to the RPA dispersion, can be qualitatively reproduced at small $k_\perp \rL$ for $\Gamma=0.25$. On the other hand, the BGK model does not provide consistent agreement with the simulations for the DSF, see Fig.~\ref{fig:dispb058}(b) and (c). While the shape of the BGK DSF in (c) is in reasonable agreement with the MD data for $\Gamma=0.25$, the peaks in (b) at a smaller wave number are not well reproduced, and the damping is overestimated. Increasing the relaxation rate in (c) to $\nu/\wpl=0.5$ (short dashed line) leads to the vanishing of the upper peak, whereas an increase of the coupling strength in the simulations appears to favor the opposite behavior. For completeness, the QLCA dispersion relation for the UH mode is also displayed in Fig.~\ref{fig:dispb058}(a) for $\Gamma=3$, even though the coupling is weaker than its anticipated range of applicability. The QLCA is in reasonable agreement with the simulations but shows a negative dispersion at small $k_\perp$ whereas the peak frequency from the simulations slightly increases with $k_\perp$ in this regime. 

None of the theories can provide a consistent explanation for the simulation results at all coupling strengths.  The RPA and BGK models contain thermal and kinetic effects, which are missing in the QLCA. On the other hand, the QLCA contains strong coupling effects via the pair distribution, which lead, e.g., to a change of the elastic constants~\cite{kaehlert2015cpp}. While the treatment of collisions in the BGK dielectric function is simplistic, the RPA and QLCA are missing these effects completely. In particular for a scenario as in Fig.~\ref{fig:dispb058}, where the plasma is (i) moderately coupled, (ii) moderately magnetized, and (iii) the upper hybrid frequency is in the direct vicinity of a Bernstein mode, all of these effects may play a role simultaneously.

\subsection{Oblique angles}
\begin{figure}
\includegraphics{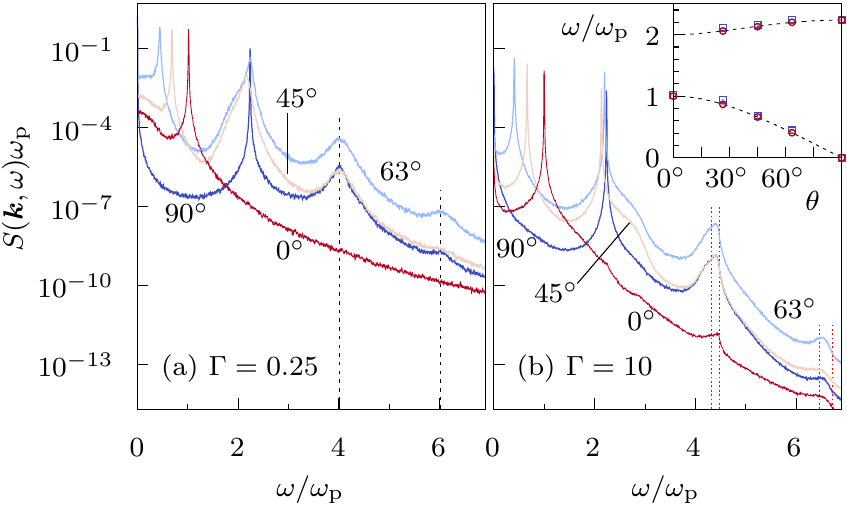}
\caption{\label{fig:pptransition}DSF of the OCP with $\beta=2$ for various angles $\theta=\angle(\vec k,\vec B)$ and two coupling strengths, as indicated in the figure. The vertical lines depict harmonics of (a) $\wc$ and (b) $\wuh$ and $\omega_\infty$ ($\omega_\infty<\wuh$). The wave number is $k_\text{min}a=0.0905$ ($0\degree$ and $90\degree$), $ka=\sqrt{2}\,k_\text{min}a$ $(45\degree)$, and $ka=\sqrt{5}\,k_\text{min}a$ ($63\degree$). The inset in (b) shows the frequencies of the two main peaks as a function of $\theta$. Squares (circles) correspond to $\Gamma=0.25$ ($\Gamma=10$). The dashed lines are the frequencies of cold fluid theory, see Ref.~\cite{kaehlert2013}.}
\end{figure}

In the following, the DSF will be inspected for oblique angles between $\vec k$ and $\vec B$. Consider first Fig.~\ref{fig:pptransition}, which shows the transition as $\theta$ is varied from $90\degree$ to $0\degree$ in the long wavelength limit. For $\Gamma=0.25$ [Fig.~\ref{fig:pptransition}(a)], the DSF at $90\degree$ is dominated by the upper hybrid mode and the second cyclotron harmonic. The third harmonic is barely visible. On the other hand, for $\theta=0\degree$, the spectrum only shows the plasmon mode at the plasma frequency. Under oblique angles, two principal collective modes can be observed. The first has a frequency below the plasma frequency while the second mode remains close to the upper hybrid frequency. In addition, the second cyclotron harmonic persists while the third becomes weak for $\theta=45\degree$.

At $\Gamma=10$ [Fig.~\ref{fig:pptransition}(b)], the low frequency part of the spectrum is very similar to the weakly coupled case in Fig.~\ref{fig:pptransition}(a). In fact, the frequencies of the two main collective modes are well described by the cold fluid result~\cite{kaehlert2013}, as shown in the inset in Fig.~\ref{fig:pptransition}(b), both for weak and strong coupling. For $\beta>1$, the frequency of the upper mode branch lies between the cyclotron frequency and the upper hybrid frequency while the lower branch has a frequency below the plasma frequency and approaches zero as $\theta\to 90\degree$. In the high frequency part of the spectrum, there now appear harmonics just below $2\,\wuh$ and $3\,\wuh$, as discussed in Sec.~\ref{sssec:HH}. This feature in the DSF persists even for $\theta=0\degree$, but with a lower intensity, as already observed in the spectrum of the longitudinal current, which is closely related to the DSF~\cite{ott2012}.

\begin{figure}
\includegraphics{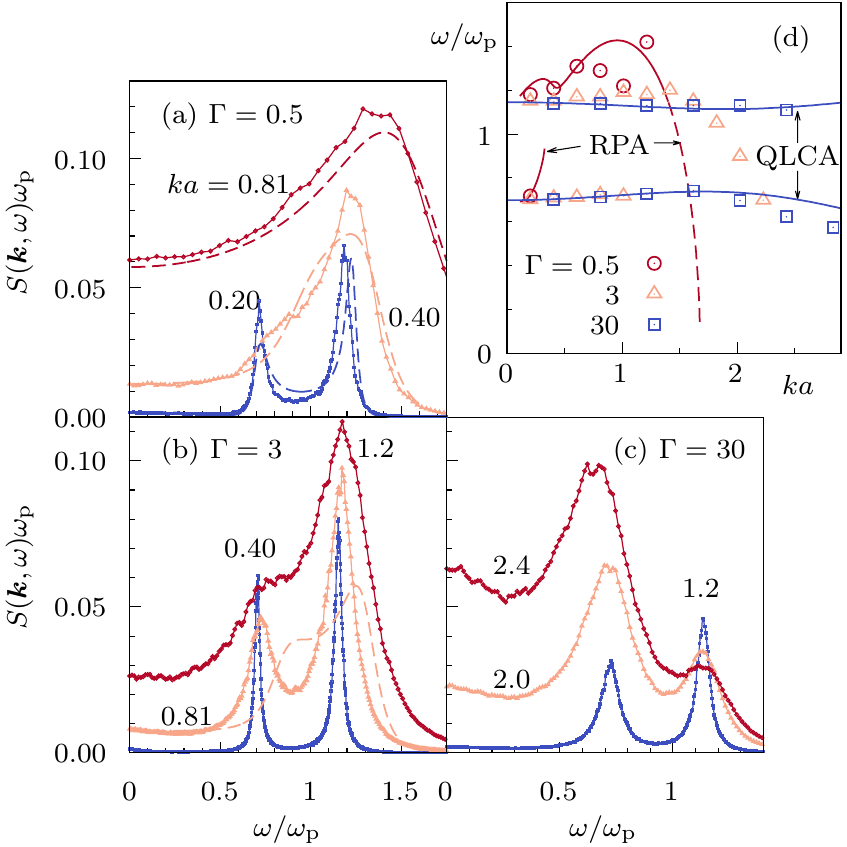}
\caption{\label{fig:dispobl}(a)-(c) DSF at an angle $\theta\approx 27\degree$ between $\vec k$ and $\vec B$ for $\beta=2/\sqrt{5}\approx 0.89$. The wave number $k=|\vec k|$ is indicated in the figure. Dashed lines in (a) and (b) correspond to the RPA. (d) Peak position of the DSF. The solid lines show the result of the RPA at $\Gamma=0.5$ (QLCA at $\Gamma=30$). In case of the RPA, the dashed line indicates the interval where the peak becomes barely visible.}
\end{figure}

The DSF and the peak positions of the two principal collective modes for $\theta\approx 27\degree$ are shown in Fig.~\ref{fig:dispobl} for $\beta\approx 0.89$. Note that, for $\beta<1$ and in the $|\vec k|\to 0$ limit, cold fluid theory predicts the frequencies in the upper and lower branch to lie between $\wpl$ and $\wuh$, and $\wc$ and $0$, respectively. As the wave number $|\vec k|$ increases at fixed $\theta$, the second peak in the moderately coupled system with $\Gamma=0.5$ becomes dominant [Fig.~\ref{fig:dispobl}(a)], and its frequency increases. Also shown is the DSF from the RPA, which is in good agreement with the simulations. At $\Gamma=3$ [Fig.~\ref{fig:dispobl}(b)], the trend remains intact, but the RPA fails to provide an adequate description of the simulations. Finally, deep in the strongly coupled regime [$\Gamma=30$, Fig.~\ref{fig:dispobl}(c)], the trend is reversed. Here, the high frequency peak gradually becomes weaker while, at the same time, the low frequency mode grows in intensity and eventually becomes dominant.

The peak positions of the main modes also show a qualitatively different behavior at weak and strong coupling, see Fig.~\ref{fig:dispobl}(d). For $\Gamma=0.5$, the frequency of the upper peak increases with $|\vec k|$, which is in qualitative agreement with the RPA. Note, however, that the determination of the frequency is rendered difficult due to the somewhat noisy data. A rapid drop of the RPA peak frequency then occurs for $ka\gtrsim 1.4$, where the peak intensity also becomes significantly weaker (indicated by the dashed line). In the intermediate coupling regime ($\Gamma=3$), the lower peak persists for a larger wave number interval and can be detected up to $ka\approx 1$. The frequency of the upper peak is now almost constant and starts to decrease for $ka\gtrsim 1.5$. For $\Gamma=30$, the lower mode frequency initially increases slightly and drops beyond $ka\gtrsim 2$. The frequency of the upper mode is almost constant. In this coupling regime, the frequencies are well described by the QLCA dispersion relation~\cite{kaehlert2013}.

\subsection{Wave vector parallel to magnetic field}
As discussed in Sec.~\ref{sec:theory}, neither the RPA nor the QLCA predict any change of the DSF or the dispersion relation of the plasmon for wave vectors strictly along the magnetic field. The DSF obtained from the simulations, however, is altered by the magnetic field.

\begin{figure}
\includegraphics{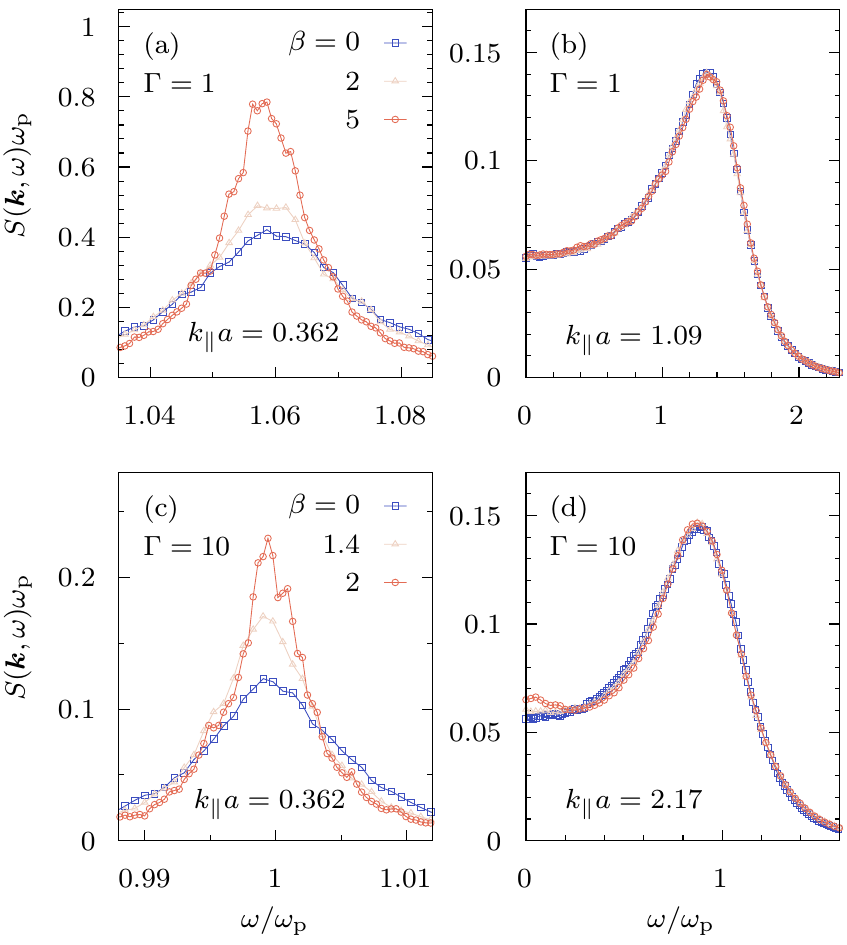}
\caption{\label{fig:plasmon}DSF for $\vec k\parallel \vec B$ at (a),(b)  $\Gamma=1$ and (c),(d) $\Gamma=10$. The wave numbers and magnetization are indicated in the figure.}
\end{figure}

The plasmon mode is shown in detail in Fig.~\ref{fig:plasmon}. The peak position is indeed largely independent of the field. However, at small wave numbers, the peak width becomes smaller as the magnetization increases, see Fig.~\ref{fig:plasmon}(a) and (c). This effect is more pronounced in the system with stronger coupling ($\Gamma=10$), where the peak height roughly doubles from $\beta=0$ to $\beta=2$. At $\Gamma=1$, a magnetization $\beta=5$ is required for a similar effect.  At larger wave numbers, the influence of the magnetic field becomes marginal, and the DSF is almost independent of $\beta$. Only for $\Gamma=10$ and $\beta=2$, the simulations show signs for the formation of a maximum at $\omega=0$.

\begin{figure}
\includegraphics{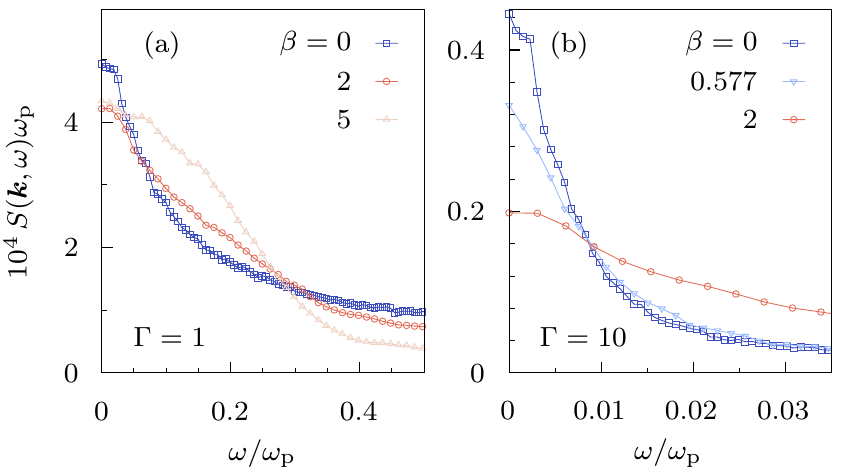}
\caption{\label{fig:zerofreq}DSF for the wave vector parallel to the magnetic field with $k_\parallel a=0.181$. The coupling and magnetization are indicated in the figure.}
\end{figure}

The low frequency part of the spectrum is investigated in Fig.~\ref{fig:zerofreq}. Note that the intensity of the DSF in this frequency range is orders of magnitude lower than for the plasmon. As opposed to the latter, an increase of the magnetization leads to a broadening of the zero frequency peak. Note, however, that the decay of the peak can become very rapid, which is best observed in the $\Gamma=1$ and $\beta=5$ case. As before, the fields required to observe these effects are smaller when the system is strongly coupled.

In dense Yukawa liquids, where the interaction is short-ranged, hydrodynamics provides an excellent description of the DSF in the small wave number and low frequency limit~\cite{mithen2011prerapid,kaehlert2020prr}. Moreover, it establishes a close link between the peak widths of the diffusive heat mode and the sound mode on the one hand, and the thermal diffusivity and the longitudinal viscosity on the other hand. In case of the OCP, however, it was found~\cite{mithen2011prerapid} that the coefficients obtained from a fit of the hydrodynamic DSF of Baus and Hansen~\cite{baus1980} to simulation data did not agree well with the transport coefficients of the OCP~\cite{daligault2006prl}. Thus, it remains unclear if a connection between the transport coefficients and the DSF can be established for the (magnetized) OCP. If so, it could provide a means to determine the latter directly from the DSF.

\section{Conclusions}\label{sec:conclusions}
In summary, the DSF of the magnetized OCP has been studied across coupling regimes. The simulations have shown that the DSF and the wave spectrum undergo major changes when the coupling regimes are crossed.

(i) In weakly coupled plasmas and for wave vectors perpendicular to the field, the DSF exhibits higher harmonics of the cyclotron frequency that can be attributed to Bernstein modes. On the other hand, (ii) in strongly coupled systems, the higher harmonics found in this regime are related to the upper hybrid mode, see also Ref.~\cite{ott2012}. Unless the plasma is strongly magnetized, $\beta>1$, the excitation of harmonics is weak in the intermediate regime around $\Gamma\approx 1-3$, where the crossover occurs. In weakly to moderately magnetized systems, when the upper hybrid frequency is close to the second cyclotron harmonic, the simulations have revealed an intricate modification of the dispersion relation for frequencies in the upper hybrid range. A spectrum with two well separated modes at weak coupling transforms into a single mode spectrum at strong coupling.

(iii) Under oblique angles, two principal collective modes are observed whose frequencies are well described by the cold fluid result in the long wavelength limit. The intensity of the peaks strongly depends on $\Gamma$ and $\beta$. For a particular case with $\beta\approx 0.89$, the upper (lower) mode becomes dominant as the wave number increases when the system is weakly (strongly) coupled. In the direction parallel to the external field, the simulations have shown that (iv) the peak position of the plasmon mode is practically unaffected by the magnetic field. On the other hand, the peaks become significantly narrower in strongly magnetized plasmas. This effect becomes marginal, however, when the wave number is increased.

The RPA dielectric function can provide a good description for the DSF and  the dispersion relation of the collective modes in weakly coupled plasmas, $\Gamma\ll 1$. However, it misses collisional damping, which can be accounted for in a Mermin type dielectric function with a particle number conserving BGK collision operator~\cite{nersisyan2011}. While it can reproduce the line broadening qualitatively, the model could be improved by including momentum and energy conservation, which is particularly important in the long wavelength limit, as has been demonstrated in unmagnetized systems~\cite{atwal2002}. In the strongly coupled regime, the QLCA provides a good description of the dispersion relation of the collective modes but misses damping effects. None of the approaches can be consistently applied in the intermediate range, $\Gamma\sim\mathcal O(1)$, or reproduce the crossover between weak and strong coupling. This calls for a unified theory that properly accounts for kinetic, collisional, and strong coupling effects at the same time.

While the OCP allows one to study in detail the simultaneous effects of strong coupling and magnetization, the physics may become considerably more complex in realistic two-component plasmas. Due to the large mass difference between electrons and ions, the former are more easily magnetized than the latter, and the plasma could be partially magnetized. Moreover, depending on the density and temperature, the electrons can be degenerate, as e.g., in warm dense matter~\cite{bonitz2020, graziani2014book}. Therefore, an extension of the present work to the magnetized uniform (quantum) electron gas at finite temperature would be of high interest, where an external magnetic field even affects the equilibrium properties of the system, in contrast to the classical OCP. This would also affect the screening of the (classical) ions by degenerate electrons~\cite{moldabekov2015pop}, which is of direct relevance for their effective coupling strength~\cite{ott2011,ott2014pop} and the ion-ion DSF~\cite{moldabekov2019}. While exact dynamical simulations for the uniform electron gas are unfeasible, ab initio data for the DSF have recently become available from quantum Monte Carlo simulations (for unmagnetized systems)~\cite{dornheim2018prl}. In the magnetized case, RPA results have been derived for the response functions and the mode spectrum~\cite{horing1965,horing2011}. Finally, it must be kept in mind that the OCP is an electrostatic model and therefore misses many of the collective modes that emerge in a full electromagnetic description of magnetized plasmas.

\begin{acknowledgments}
The molecular dynamics simulations were performed at the Norddeutscher Verbund für Hoch- und Höchstleistungsrechnen (HLRN) under grant shp00026.
\end{acknowledgments}

\appendix
\section{Moments of the DSF}\label{sec:appendix}
The frequency moments of the DSF,
\begin{align}\label{eqn:defmom}
    \langle \omega^m \rangle =\int_{-\infty}^\infty\omega^m\, S(\vec k,\omega)\, d\omega,
\end{align}
are determined by the static properties of the system. Recalling the derivation of the moments~\cite{balucani1995book,hansen2006}, one observes that, in case of the magnetized OCP, the zeroth and the second moment are unaffected by the magnetic field. They are given by
$\langle \omega^0\rangle=S(k)$ and $\langle \omega^2\rangle/\wpl^2=(ka)^2/(3\,\Gamma)$, respectively. Here, $S(k)$ is the static structure factor. For the fourth moment, one finds an additional contribution $\sim k_\perp^2 \beta^2/k^2$ from the Lorentz force,
\begin{align}\label{eqn:mom4}
    \frac{\langle \omega^4\rangle }{\wpl^4}=\frac{\langle \omega^2\rangle}{\wpl^2}\left[\frac{k_\perp^2}{k^2}\beta^2+ \frac{(ka)^2}{\Gamma}+1-2\,I(k) \right],
\end{align}
where the quantity $I(k)$~\cite{mithen2012pre} also appears in the QLCA, $I(k)=-D_L(k)/(2\,\wpl^2)=D_T(k)/\wpl^2$, see Eq.~\eqref{eqn:DT} and below.

%

%\bibliography{references}

\end{document}